\documentclass[12pt]{article}
\usepackage{subcaption}
\usepackage{graphicx}
\usepackage{lineno}
\usepackage{units}
\usepackage{atlasphysics}
\usepackage{verbatim}

\usepackage{wrapfig}

\def\bbar {\ensuremath{{\overline b}}}

\def\bbbar {\ensuremath{{b\bbar}}}

\def\fb {\ensuremath{{\mathrm{fb}}}}

\newcommand\pubdate{\today}

\def\Title#1{\begin{center} {\Large #1 } \end{center}}
\def\Author#1{\begin{center}{ \sc #1} \end{center}}
\def\Address#1{\begin{center}{ \it #1} \end{center}}

\newcommand\pubblock{\rightline{\begin{tabular}{l}  \\ 
         \pubdate  \end{tabular}}}
\newenvironment{Abstract}{\begin{quotation}  }{\end{quotation}}
\newenvironment{Presented}{\begin{quotation} \begin{center} 
             PRESENTED AT\end{center}\bigskip 
      \begin{center}\begin{large}}{\end{large}\end{center} \end{quotation}}

\begin{document}


\begin{titlepage}
 \pubblock
\vfill
\Title{Search for rare decays and lepton-flavor-violating decays \\
of the Higgs boson at the ATLAS experiment}
\vfill
\Author{Pawel Bruckman de Renstrom \\
         on behalf of the ATLAS Collaboration}
\Address{Institute of Nuclear Physics PAN, Cracow, Poland}
\vfill
\begin{Abstract}
  The Standard Model predicts several rare Higgs boson decay channels, among which are the decays to a Z boson and a photon, to a low-mass lepton pair and a photon, and to a meson and a photon.
  The observation of some of these decays could open the possibility of studying the CP and coupling properties of the Higgs boson in a complementary way to other analyses.
  In addition, lepton-flavor-violating decays of the observed Higgs boson are searched for, where an observation would be a clear sign of physics effects beyond the Standard Model.
  These proceedings present selected recent results for such decays based on proton–proton collision data at $13$~TeV collected by the ATLAS experiment in Run 2 of the LHC.
\end{Abstract}
\vfill
\begin{Presented}
DIS2023: XXX International Workshop on Deep-Inelastic Scattering and
Related Subjects, \\
Michigan State University, USA, 27-31 March 2023 \\
     \includegraphics[width=9cm]{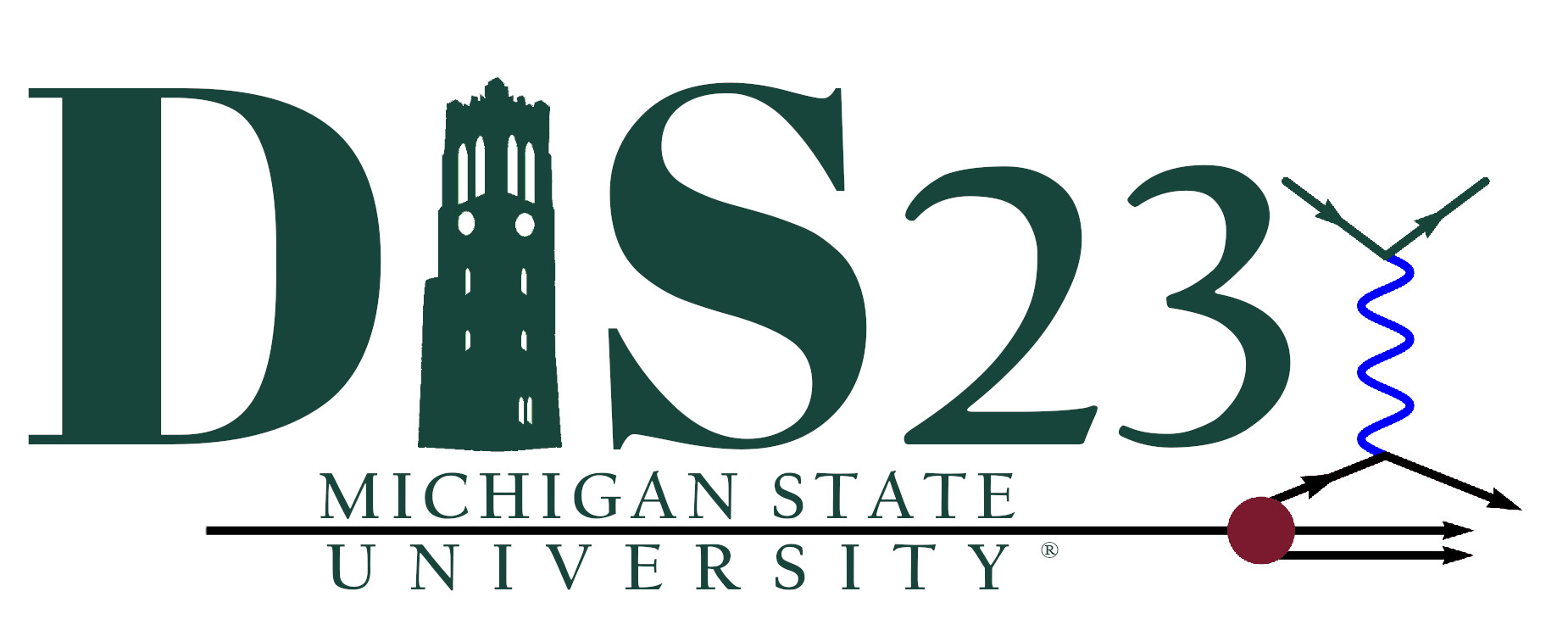}
\end{Presented}
\vfill
\small{Copyright [2023] CERN for the benefit of the ATLAS Collaboration. CC-BY-4.0 license.}
\end{titlepage}

\section{Introduction}

Since the discovery of the Higgs boson by the CMS and ATLAS Collaborations in 2012, its properties have been measured to increasing precision. 
So far, an excellent agreement with the predictions for a Standard Model (SM) Higgs boson is observed.
However, the SM, while highly successful, is not considered to be a complete theory as it is not capable of explaining 
some of the phenomena seen in nature.
Studying Higgs decays which are highly suppressed or forbidden in the SM is particularly attractive to probe for the smoking gun of phenomena from beyond the SM.
In the following, three categories of search analyses are reported.
The first concerns with Higgs decays occurring via loop diagrams or suppressed by the small Yukawa coupling constant to the first or the second generation fermions.
The second one are searches for lepton flavour violating (LFV) decays of the Higgs boson.
The third category are Higgs decays to the first or the second generation leptons. 
The proceedings summarizes selected recent search results, obtained using proton–proton ($pp$) collision data at $13$~TeV collected 
by the ATLAS experiment~\cite{atlas} in Run 2 of the LHC.
All presented results concern rare processes and are statistically limited.
Hence, no discussion of systematic uncertainties is included in the following. 

\section{Search for rare loop-induced or Yukawa-suppressed Higgs decays}
\begin{figure}[t]
\centering
\begin{subfigure}[t]{0.47\textwidth}
    \begin{center}
    \includegraphics[width=0.87\textwidth]{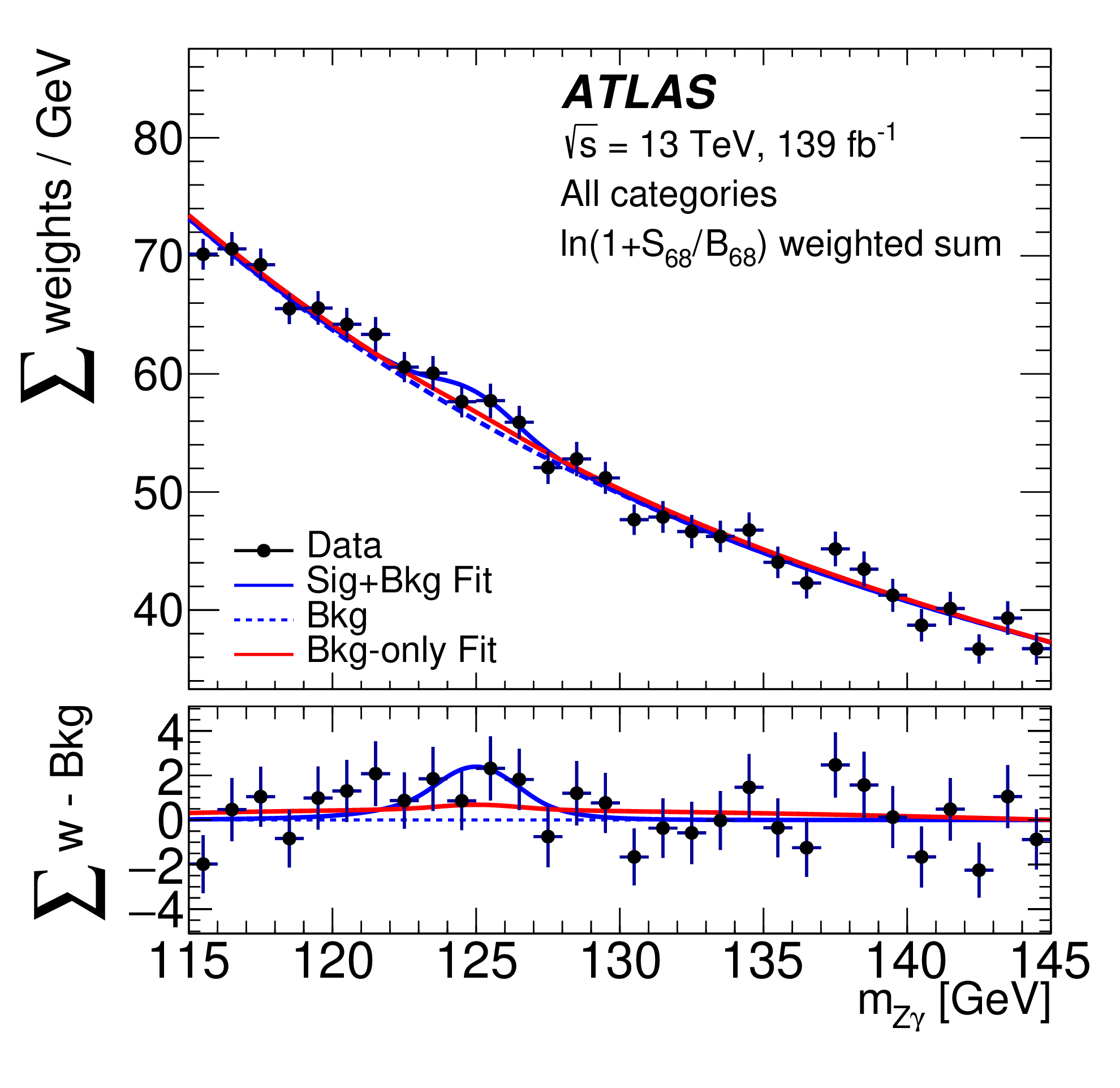}
    \end{center}
    \vspace*{-0.5cm}
    \caption{
      $H \rightarrow Z\gamma $ selection
      \cite{Zgamma}
      \label{fig:Zy}
      }
\end{subfigure}
\hfill
\begin{subfigure}[t]{0.47\textwidth}
    \begin{center}
    \includegraphics[width=0.85\textwidth]{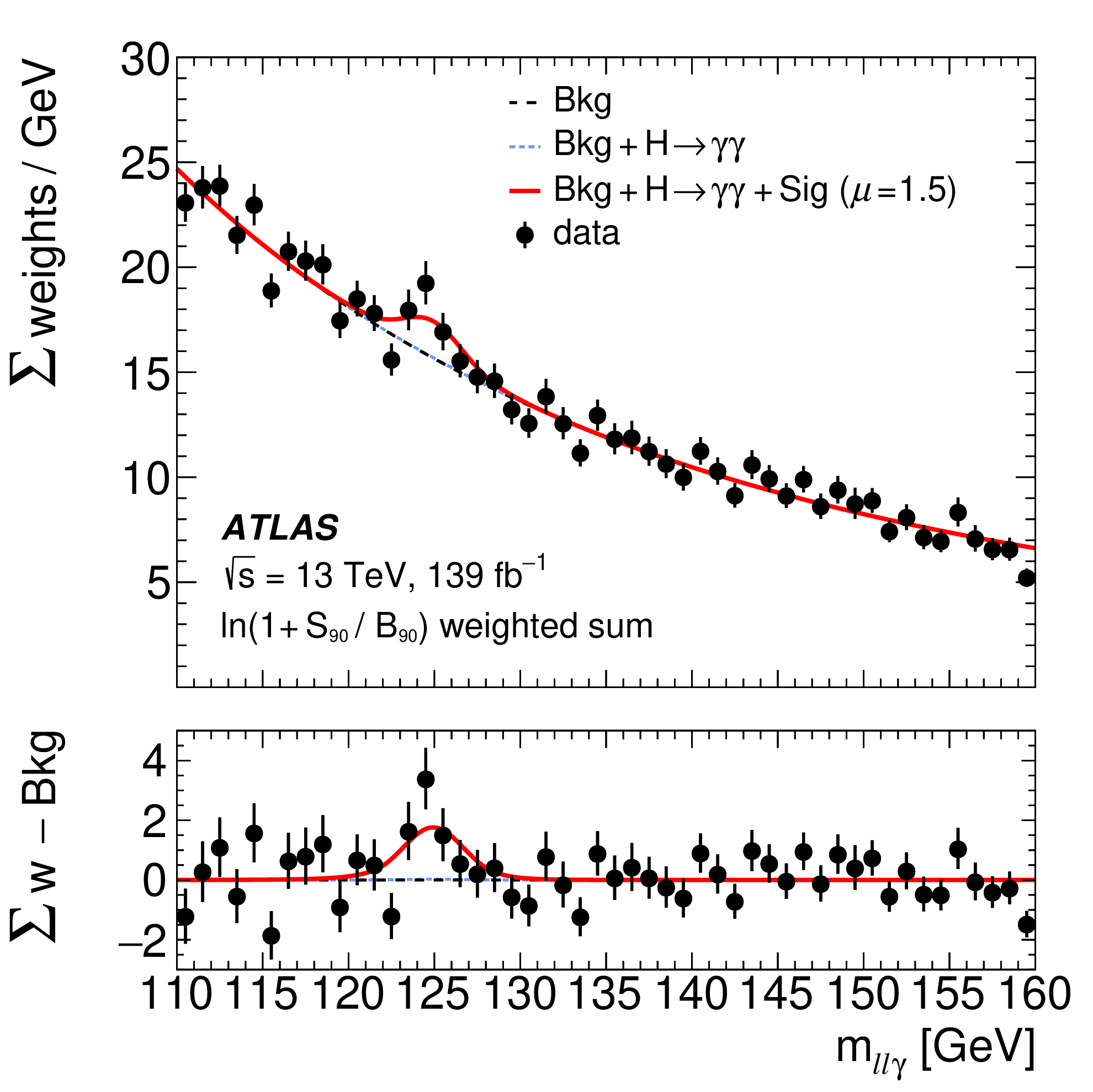}
    \end{center}
    \vspace*{-0.5cm}
    \caption{
      $H \rightarrow \ell \ell \gamma $ selection 
      \cite{llgamma}
      \label{fig:yy}
      }
\end{subfigure}
\caption{
  Invariant mass distribution of events satisfying signal selection criteria in data and fit results while fitting all signal categories simultaneously.
  Bottom panels show the residuals of the data with respect to the non-resonant background component of the signal plus background fit.
}
\end{figure}

Higgs decays to quarkonia or light mesons with an emission of a prompt photon occur via loop diagrams or direct diagrams suppressed by the small Yukawa coupling constant to the first or the second generation fermions.
The branching fraction expected in the SM may be additionally suppressed due to strong destructive interference of the loop and the direct decay diagrams.
Modifications to either Yukawa couplings or additional contributions to the loop diagrams could substantially enhance the branching. \newline
\textbf{\boldmath$H \rightarrow Z\gamma$ with leptonic \boldmath$Z$ decays} \newline
The SM Higgs boson can decay into $Z\gamma$ through loop diagrams and the branching ratio is predicted to be $B(H \rightarrow Z\gamma) = (1.54 \pm 0.09) \times 10^{-3}$ at $m_H = 125.09 \GeV $~\cite{handbook}.
It can differ from the SM value for several scenarios beyond the SM, for example, if the Higgs boson was a neutral scalar of different origin, or a composite state.
Different branching ratios are also expected for models with additional colourless charged scalars, leptons or vector bosons that couple to the Higgs boson, due to their contributions via loop corrections.
 A search for $Z\gamma$ decays of the SM Higgs boson in 139~fb$^{-1}$ of $pp$ collisions at $\sqrt{s}=13 \TeV $ is performed in the leptonic $Z$ decay channel, where lepton accounts for either an electron or a muon~\cite{Zgamma}. The invariant mass $m_{Z\gamma}$ is reconstructed using kinematical fit with the $Z$ mass constraint on the emerging lepton pair.
 The resulting mass spectrum is symultaeously fitted in six mutually exclusive signal regions.
 The weighted sum of all the region together with the fit results is shown in Fig.~\ref{fig:Zy}.
The observed data are consistent with the expected background
with the $p$-value of 1.3\%, while the expected $p$-value in the presence of the SM Higgs boson is 12.3\%, corresponding to significances of 2.2 and 1.2~$\sigma$, respectively.
The observed 95\%\ confidence level (CL) upper limit on the $\sigma(pp \rightarrow H) \times B(H \rightarrow Z\gamma)$ is 3.6 times the SM prediction for the Higgs boson mass of 125.09 \GeV .
The expected limit on $\sigma(pp \rightarrow H) \times B(H \rightarrow Z\gamma)$ assuming either no Higgs boson decay into $Z\gamma$ or the presence of the SM Higgs boson decay is 1.7 and 2.6 times the SM prediction, respectively.
The best-fit value for the signal yield normalised to the SM prediction is $2.0^{+1.0}_{-0.9}$ where the statistical component of the uncertainty is dominant.

\medskip
\noindent
\textbf{\boldmath$H \rightarrow \gamma^* \gamma$ with \boldmath$\gamma^*$ decaying into a low mass lepton pair} \newline
The analysis targets quasi the same final state as the one discussed above, but the requirement of low invariant mass of the lepton pair ($m_{\ell \ell}<30 \GeV$) makes this signal region completely dominated by the decay through a $\gamma^*$~\cite{llgamma}.
With the requirement of low $\ell \ell$ mass and high $p_{\rm T}$ electron clusters can overlap, hence the analysis introduces an additional signal category of merged electrons.
Consequently, a dedicated reconstruction of merged electron pairs, based on machine learning methods is put in place.
The weighted sum of the resulting 9 mutually exclusive signal regions together with the fit result is shown in Fig.~\ref{fig:yy}.
For a Higgs boson with a mass of 125.09 GeV and $m_{\ell \ell}<30 \GeV$, evidence for the $H \rightarrow \ell \ell \gamma$ process is found with the significance of $3.2 \sigma$ over the background-only hypothesis, compared to an expected significance of $2.1 \sigma$.
The best-fit value of the signal-strength parameter, defined as the ratio of the observed signal yield to the signal yield expected in the SM, is $\mu = 1.5\pm 0.5$.
The Higgs boson production cross-section times the $H \rightarrow \ell \ell \gamma$ branching ratio for $m_{\ell \ell}<30 \GeV$ is determined to be $8.7^{+2.8}_{-2.7}\ \fb$.
This result constitutes the first evidence for the decay of the Higgs boson into a pair of leptons
and a photon, an important step towards probing Higgs boson couplings in this rare decay channel.

\begin{figure}[h]
\centering
\begin{subfigure}[t]{0.49\textwidth}
  \begin{center}
    \includegraphics[width=0.85\textwidth]{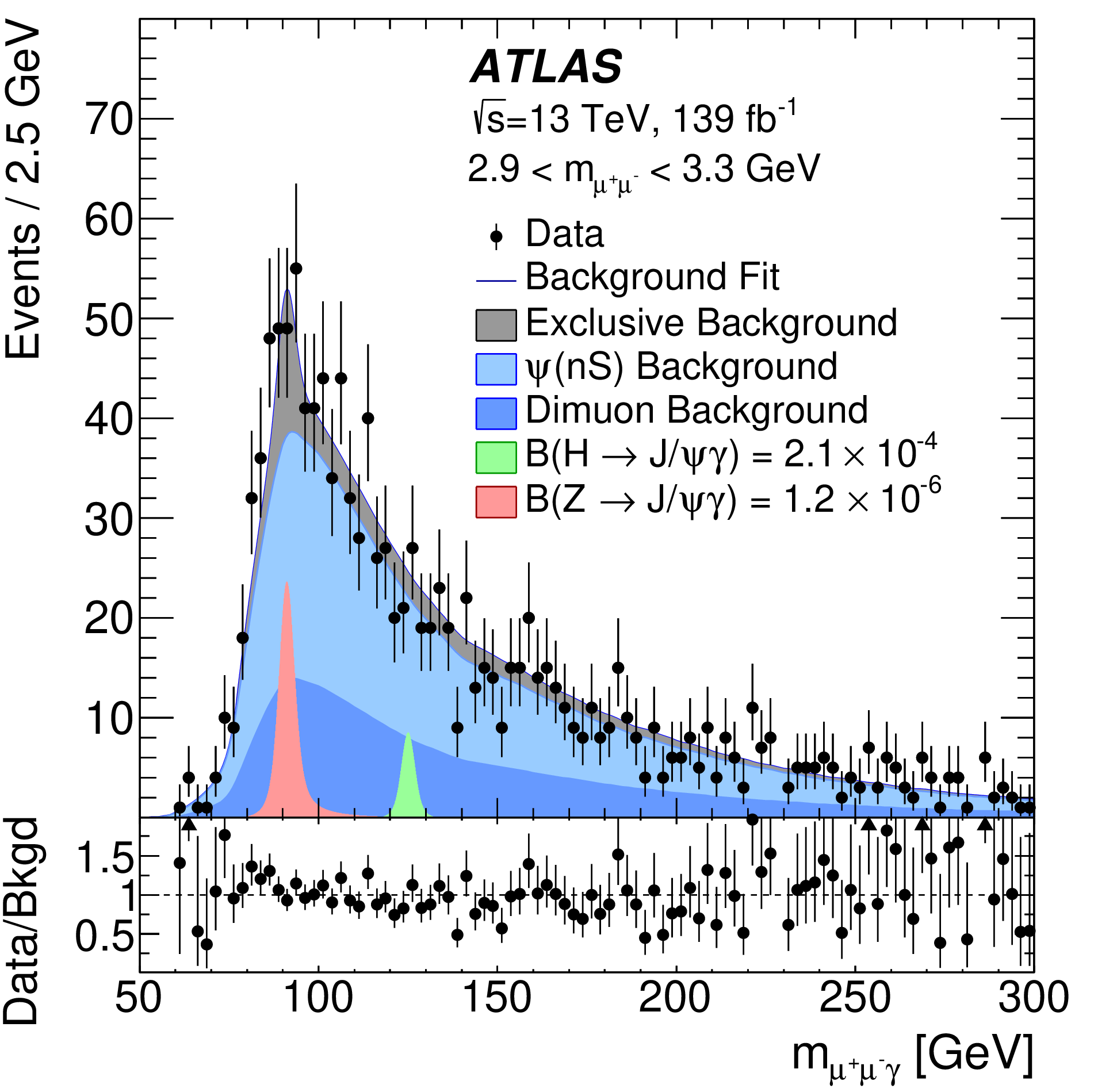}
  \end{center}
    \vspace*{-0.5cm}
    \caption{ }
\end{subfigure}
\hfill
\begin{subfigure}[t]{0.49\textwidth}
  \begin{center}
    \includegraphics[width=0.85\textwidth]{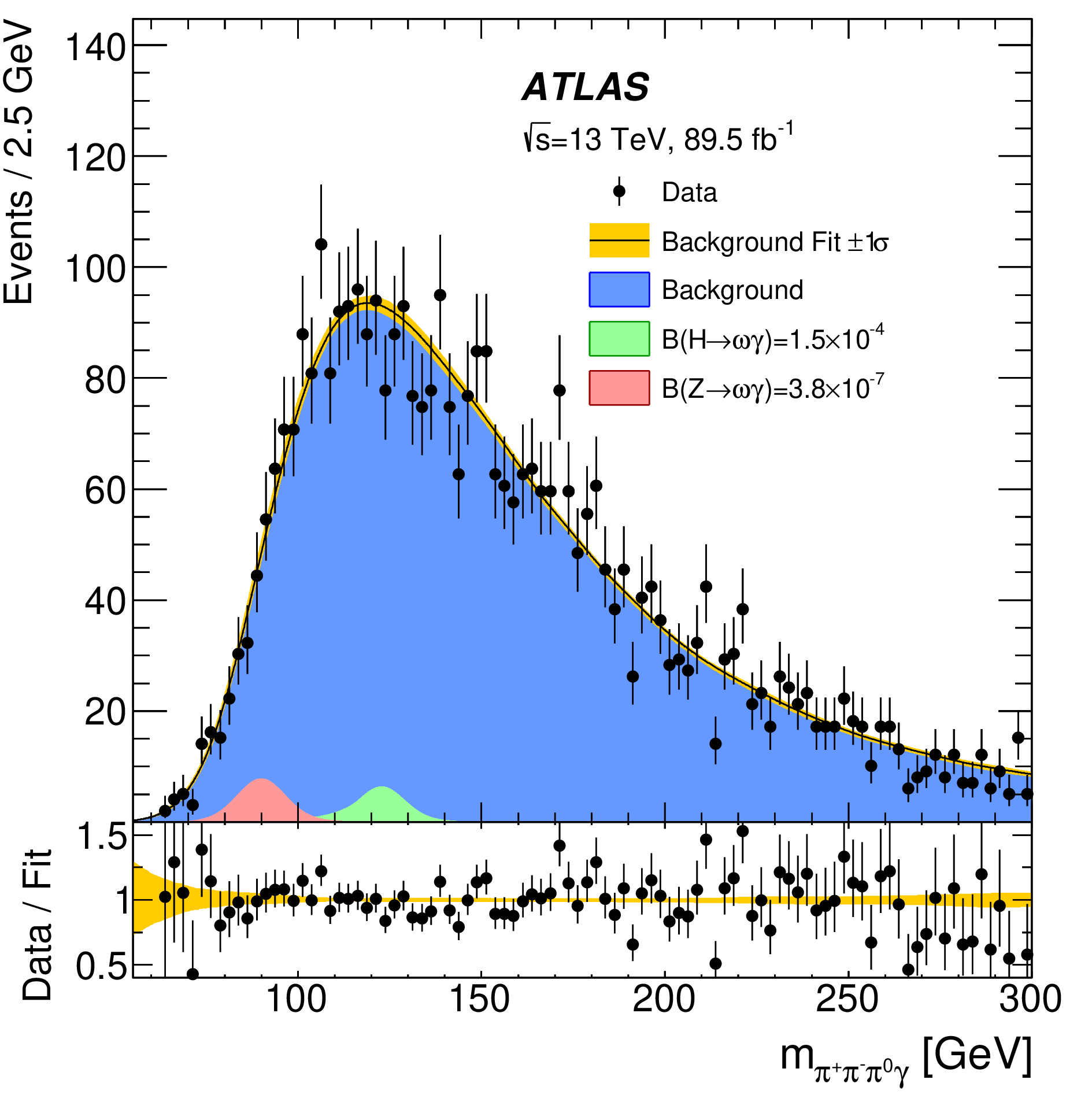}
  \end{center}
    \vspace*{-0.5cm}
    \caption{ }
\end{subfigure}
\caption{(a) Reconstructed invariant mass $m_{\mu^+ \mu^- \gamma}$ for the $J\psi \gamma$~\cite{quarkonia} and (b) reconstructed invariant mass $m_{\pi^+ \pi^- \pi^0 \gamma}$ for the $\omega \gamma$ search~\cite{lightmesons}.  The branching fraction of each of the signals is set to the observed 95\%\ CL upper limit. 
  \label{fig:excl}
}
\end{figure}

\medskip
\noindent
\textbf{ Higgs decays to exclusivly reconstructed vector quarkonia or light mesons and a photon} \newline
Due to highly destrucive interference between the indirect and direct Feynman diagrams, Higgs boson decays in the charmonium sector, $H \rightarrow J/\psi \gamma$ and $H \rightarrow \psi (2S) \gamma$, offer an opportunity to access both the magnitude and the sign of the charm-quark  Yukawa coupling; the corresponding decays in the bottomonium sector, $H \rightarrow \Upsilon (1S, 2S, 3S) \gamma$, can provide information about the real and imaginary parts of the bottom-quark coupling to the Higgs boson.
Exclusive decays to $\omega \gamma$ and $K^* \gamma $ can probe the flavour-conserving coupling of the Higgs boson to up and down quarks, and the flavour-violating coupling of the Higgs boson to down and strange quarks, respectively.
The theoretical predictions for the corresponding branching fractions are tiny and range from $\mathcal{O}(10^{-6})$ for $J/\psi \gamma$ and  $\omega \gamma$ final states through $\mathcal{O}(10^{-9} - 10^{-8})$ for $\Upsilon \gamma$ and a rough estimate of $\ll 10^{-11}$ for $K^* \gamma $.
Searches for the above decays are performed using $\mu^+ \mu^-$ decays of the vector quarkonia~\cite{quarkonia}, while the decay $\omega \rightarrow \pi^+ \pi^- \pi^0 $ is used to reconstruct the $\omega$ meson, and the decay $K^* \rightarrow K^+ \pi^-$ is used to reconstruct the $K^*$ meson~\cite{lightmesons}.
The non-trivial background shape is estimated using the correlated sampling method~\cite{sampling} and normalised from the fit to data, as can be seen in Fig.~\ref{fig:excl} for $J/\psi \gamma$ (a) and $\omega \gamma$ (b).
  The observed data are compatible with the expected backgrounds.
The 95\%\ CLs upper limits on the branching fractions of
the Higgs boson decays into $J/\psi \gamma$, $\psi (2S) \gamma$, and $\Upsilon (1S, 2S, 3S) \gamma$ are found to be $2.1 \times 10^{-4}$, $10.9 \times 10^{-4}$, and $(2.6, 4.4, 3.5) \times 10^{-4}$, respectively, assuming SM production of the Higgs boson.
The 95\%\ CL upper limits on the branching fractions to light mesons and a photon are
$\mathcal{B}(H \rightarrow \omega \gamma) < 1.5 \times 10^{-4}$ and $\mathcal{B}(H \rightarrow K^* \gamma) < 8.9 \times 10^{-5}$.
An earlier analysis based on partial Run 2 ATLAS data extracted limits for the Higgs boson decays into $\rho \gamma$ and $\phi \gamma$ exclusive decays~\cite{rhophi}.

\section{Searches for lepton flavour violation in the Higgs sector}
\begin{figure}[h]
\centering
\begin{subfigure}[t]{0.55\textwidth}
   \begin{center}
    \includegraphics[width=0.93\textwidth]{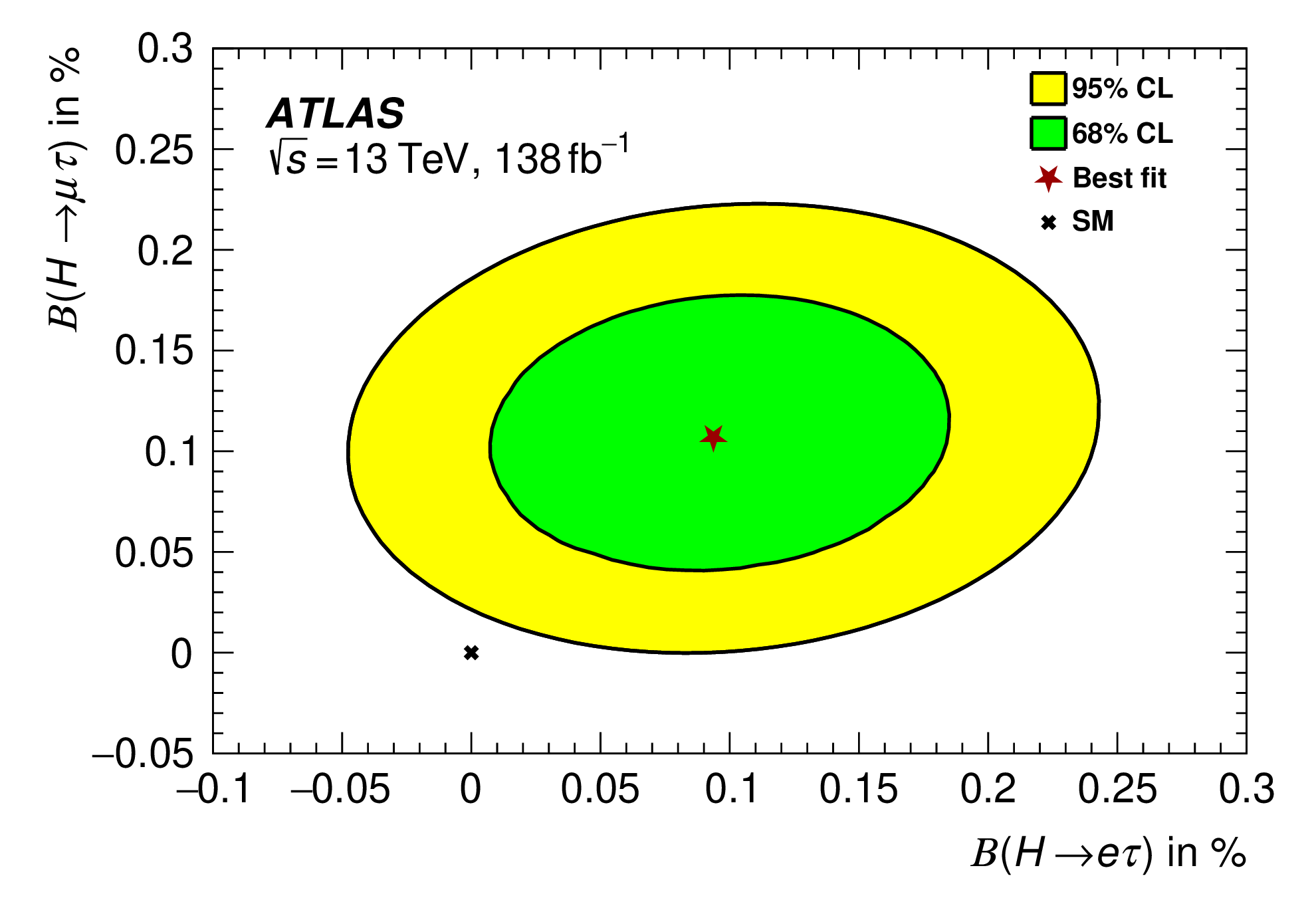}
   \end{center}
    \vspace*{-0.5cm}
    \caption{
      \label{fig:contour}
      }
\end{subfigure}
\hfill
\begin{subfigure}[t]{0.4\textwidth}
  \begin{center}
    \includegraphics[width=0.90\textwidth,trim={0 0.01cm 0 -0.5cm},clip]{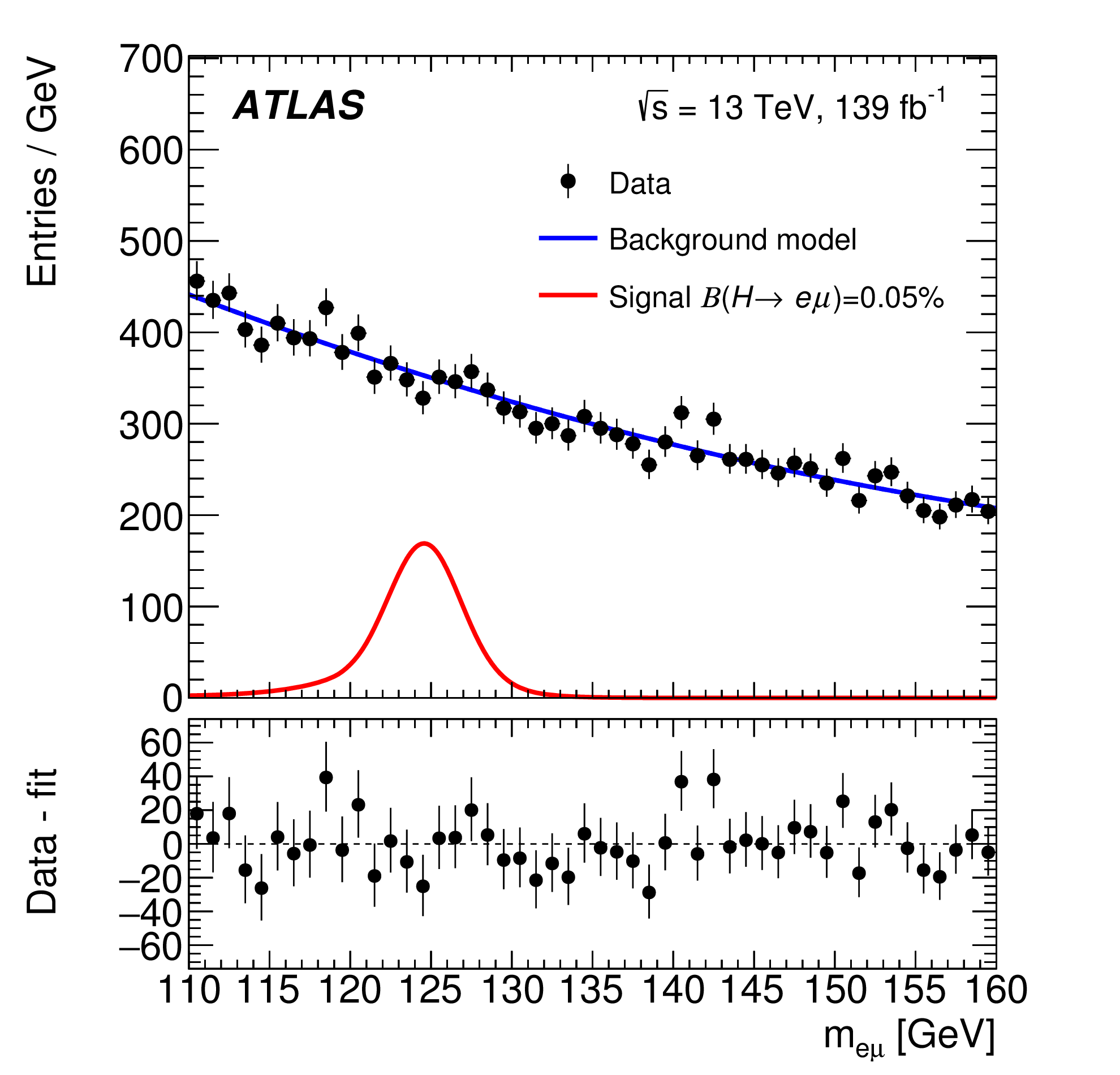}
   \end{center}
    \vspace*{-0.5cm}
    \caption{
      \label{fig:emu}
      }
\end{subfigure}
\hfill
\caption{
  (a) Best-fit value (red star) of the branching ratios $\mathcal{B}(H \rightarrow e \tau)$ and $\mathcal{B}(H \rightarrow \mu \tau)$, given in \% , and likelihood contours at 68\%\ and 95\%\ CL obtained from the simultaneous fit of $H \rightarrow e \tau$ and $H \rightarrow \mu \tau$ signals based on the MC-template method, compared with the SM expectation (black cross)~\cite{LFV_ltau}.
   (b) Simultaneous background-only fit to $e \mu$ data in all signal region cetegories.
    The signal parameterisation with branching fraction set to $\mathcal{B}(H \rightarrow e \mu) = 0.05\% $ is also shown (red line)~\cite{Heeem}.
}
\end{figure}

Neutrino oscillations prove that lepton flavour is not an exact symmetry of nature. Higgs sector may be responsible. LFV naturally occurs in more than one Higgs dublet models, composite Higgs or Randall-Sundrum warped extra dimensions.
The searches reported below allow to put direct limits on the off-diagonal elements ($Y_{\tau\mu},\ Y_{\tau e},\ Y_{e\mu}$) of the Yukawa coupling matrix without making any assumptions about diagonal couplings. \newline
\textbf{Search for the Higgs boson decay to either a \boldmath$e \tau$ or a \boldmath$\mu \tau$ pair} \newline
The direct search for lepton flavour violation in Higgs boson decays, $H \rightarrow e\tau$ and $H \rightarrow \mu\tau$ is performed using both the leptonic and hadronic $\tau$ decay channels~\cite{LFV_ltau}.
For leptonic tau decays only combinations with the different flavour lepton are considered.
Two background estimation techniques are employed: the MC-template method, based on data-corrected
simulation samples, and the Symmetry method, based on exploiting the symmetry between electrons and muons in the SM backgrounds.
No significant excess of events is observed and the results are interpreted as upper limits on lepton-flavour-violating branching ratios of the Higgs boson.
The observed (expected) upper limits set on the branching ratios at 95\%\ CL, $\mathcal{B}(H \rightarrow e\tau) < 0.20\%\ (0.12\% )$ and $\mathcal{B}(H \rightarrow \mu\tau) < 0.18\%\ (0.09\% )$, are obtained with the MC-template method from a simultaneous measurement of potential $H \rightarrow e\tau$ and $H \rightarrow \mu\tau$ signals.
The corresponding likelihood contour is shown in Fig.~\ref{fig:contour}, displaying a mild $2.1 \sigma$ tension with the SM expectation.  
The best-fit branching ratio difference, $\mathcal{B}(H \rightarrow \mu\tau) - \mathcal{B}(H \rightarrow e\tau)$, measured with the Symmetry method in the channel where the $\tau$-lepton decays to leptons, is $(0.25\pm 0.10)\% $, compatible with a value of zero within $2.5\sigma$.

\medskip
\noindent
\textbf{Search for the Higgs boson decay to a \boldmath$e \mu$ pair} \newline 
The measurement relies on simultaneous fit of the narrow resonance on top of the continuum backgrond in the invariant mass distributions reconstructed in multiple categories of signal regions~\cite{Heeem}. The categorisation is based on production mechanism, centrality and the transverse momentum of the lepton pair.
Main backrounds include Drell-Yan $\tau\tau$ pair production, $t\bar{t}$, di-boson and misindentified lepton events.
The combined result of the background-only fit to all signal cetegories is shown in Fig.~\ref{fig:emu}. 
No evidence of the decay $H \rightarrow e \mu$ is observed, with the best-fit value of the branching fraction of $(0.4 \pm 2.9 ({\rm stat.}) \pm 0.3({\rm syst.})) \times 10^{-5}$.
The observed (expected) upper limit at the 95\%\ CL is found to be $6.2 \times 10^{-5} \ \ (5.9 \times 10^{-5})$.
\begin{figure}[t]
\centering
\begin{subfigure}[t]{0.4\textwidth}
   \begin{center}
    \includegraphics[width=0.90\textwidth]{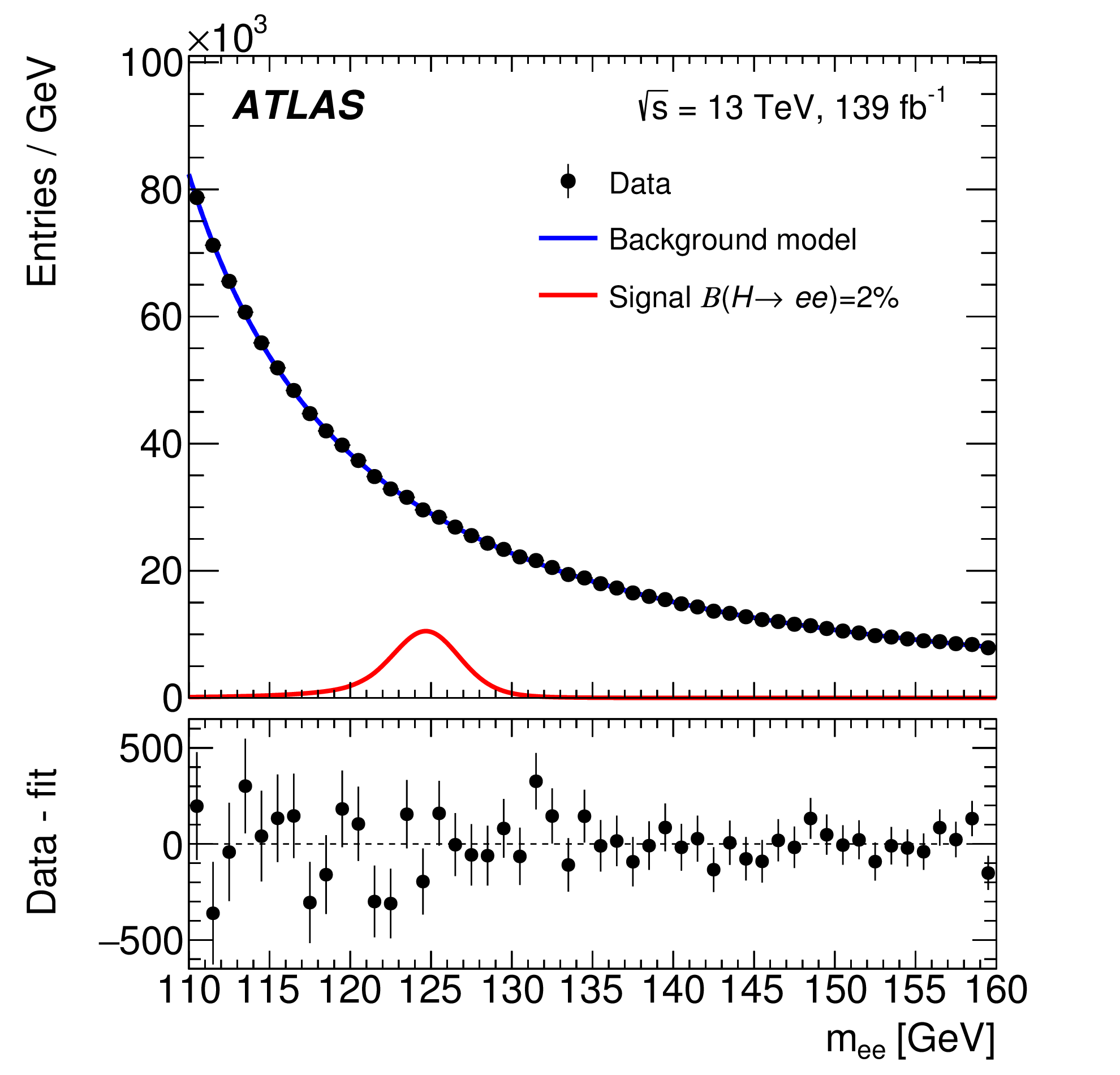}
   \end{center}
   \vspace*{-0.5cm}
   \caption{
      \label{fig:ee}
      }
\end{subfigure}
\hfill
\begin{subfigure}[t]{0.55\textwidth}
   \begin{center}
     \includegraphics[width=0.90\textwidth]{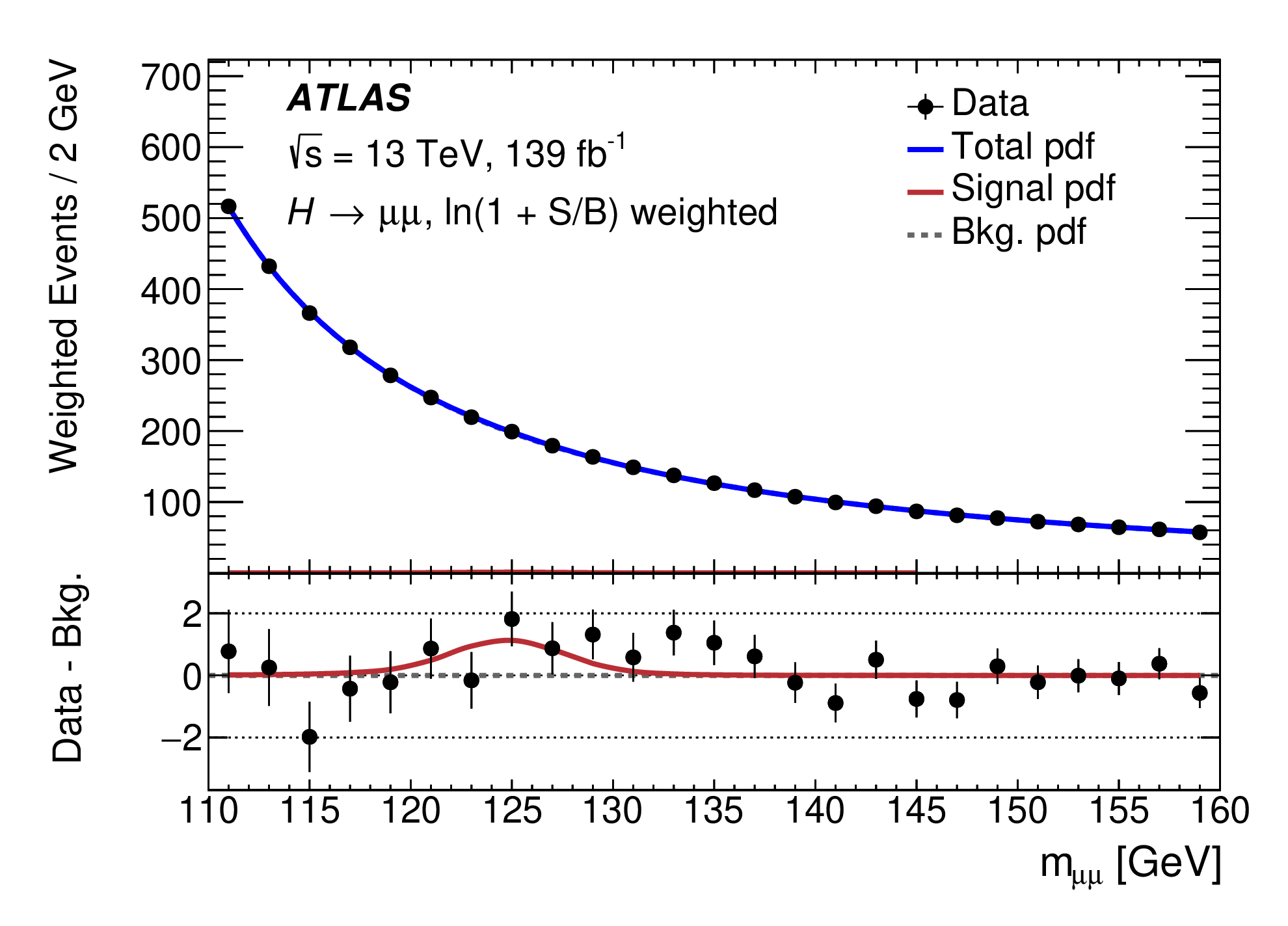}
   \end{center}
   \vspace*{-0.5cm}
   \caption{
      \label{fig:mumu}
      }
\end{subfigure}
\caption{
   (a) Simultaneous background-only fit to $e e$ data in all signal region cetegories.
  The signal parameterisation with branching fraction set to $\mathcal{B}(H \rightarrow e e) = 2\% $ is also shown (red line)~\cite{Heeem}.
   (b) Simultaneous background-only fit to $\mu \mu$ data in all signal region cetegories weighted by $\log (1 + S/B)$.
     The lower panel compares the fitted signal pdf, normalised to the signal best-fit value, to the difference between the data and the background model~\cite{Hmumu}.
}
\end{figure}
%
\section{Higgs boson decays to 1st and 2nd generation leptons}
SM predicts $\mathcal{B}(H \rightarrow e e)_{\rm SM} \sim 5 \times 10^{-9}$ and  $\mathcal{B}(H \rightarrow \mu \mu)_{\rm SM} \sim 2 \times 10^{-4}$.
Experimental verification of these predictions is important but challenging. At the same time, any significant enhancement of the branching fraction would point to phenomena from beyond the SM. \newline
\textbf{Search for the Higgs boson decay to a \boldmath$e e$ pair} \newline
The same analyis~\cite{Heeem} described above is also utilised to look for Higgs decays to opositely signed electron pairs.
Here, background is substantially larger and dominated by the Drell-Yan $ee$ production.
No evidence of the decay $H \rightarrow e e$  is observed.
The best-fit value of the branching fraction is $(0.0\pm 1.7 ({\rm stat.}) \pm 0.6({\rm syst.})) \times 10^{-4}$.
The observed (expected) upper limit on the branching fraction at the 95\%\ CL, is found to be $3.6 \times 10^{-4} \ \ (3.5 \times 10^{-4})$.

\medskip
\noindent
\textbf{Search for the Higgs boson decay to a \boldmath$\mu \mu$ pair} \newline 
A similar analysis is performed in search for muonic Higgs decays~\cite{Hmumu}.
It defines 20 mutually exclusive signal regions based on the production mechanism and S/B ratio. The clasification is done by boosted decision tree classifiers constructed for each of the four production mechanisms (ggF, VBF, VH and ttH). A fit to the reconstructed invariant mass of the muon pair is performed in all signal categories simultaneously. The invariant mass distribution weighted sum in all signal regions together with the fitted signal is shown on Fig.~\ref{fig:mumu}.
The observed (expected) significance over the background-only hypothesis for a Higgs boson with a mass of 125.09 GeV is $2.0 \sigma \ \ (1.7 \sigma)$. The observed upper limit on the cross section times branching ratio for $pp \rightarrow H \rightarrow \mu\mu$ is $2.2$ times the SM prediction at 95\%\ CL, while the expected limit on the $H \rightarrow \mu\mu$ signal assuming the absence (presence) of a SM signal is $1.1 \ \ (2.0)$. The best-fit value of the signal strength parameter, defined as the ratio of the observed signal yield to the one expected in the SM, is $\mu = 1.2 \pm 0.6$.

\section{Concluding remarks}
Searches for strongly suppressed or forbidden decays of the Higgs boson offer an atractive probe of phenomena from beyond the SM.
A wide range of such searches have been  performed by the ATLAS Colllaboration.
The full review is beyond the scope of this contribution.
Instead, the reader is encouraged to consult the ATLAS public results webpage~\cite{atlaswebpage}.
So far, no significant deviation from the SM predictions have been observed.
The sensitivity of the Higgs decay to a pair of muons is currently at the virge of an evidence.
Observation of the muonic Higgs decays as well as confirmation or disproving of the currently seen departures fro SM expectations (e.g. the LFV Higgs decay in the $\ell \tau$ channel) will require considerably more data to come with the ongoing LHC Run~3 and subsequently with the high luminosity phase of the LHC.

\bigskip
\noindent
\textbf{Acknowledgements} This work is supported in part by the Polish Ministry of Education and Science project no. 2022/WK/08.


\end{document}